# The Physics of baking good Pizza

*Andrey Varlamov[1], Andreas Glatz[2,3], Sergio Grasso[4]*

[1]CNR-SPIN, Viale del Politecnico 1, I-00133, Rome, Italy
[2]Materials Science Division, Argonne National Laboratory, 9700 South Cass Avenue, Argonne, Illinois 60439, USA
[3]Department of Physics, Northern Illinois University, DeKalb, Illinois 60115, USA
[4]Food Anthropologist, Rome, Italy

**Abstract:** Physical principles are involved in almost any aspect of cooking. Here we analyze the specific process of baking pizzas, deriving in simple terms the baking times for two different situations: For a brick oven in a pizzeria and a modern metallic oven at home. Our study is based on basic thermodynamic principles relevant to the cooking process and is accessible to undergraduate students. We start with a historical overview of the development and art of pizza baking, illustrate the underlying physics by some simple common examples, and then apply them in detail to the example of baking pizza.

**Historical background*:*

The history of pizza is crowded with tales, legends and anecdotes. Notwithstanding that, Italians are trusted as the inventors of this humble, delicious and "universal" flatbread. The precursors of pizza are the Neolithic unleavened flatbreads baked on fire-heated rocks and made from coarse grains flour (spelt, barley, emmer), developed autonomously in several areas from China to the Americas.

The Italian word "pizza" first appeared on a Latin parchment (Codex Diplomaticus Cajtanus) reporting a list of donations due by a domain tenant to the bishop of Gaeta (Naples). The document dated 997 AD fixes a supply of *duodecim pizze* ("twelve pizzas") every Christmas Day and Easter Sunday.

Etymologists debate on different origins of the word "pizza": from the Byzantine Greek πίτα=bread, cake, pie, pitta (attested in 1108); from Greek πηκτή=congealed; from Latin *picta*=painted, decorated, or *pinsere*=flatten, enlarge.

Since the 8th century BC, the Greek colonies in southern Italy (Napoli itself was founded around 600 B.C. as a Greek town), produced the *Plakuntos,* a flat baked grain-based sourdough covered with oil, garlic, onion, herbs, occasionally minced meat or small fishes, almost similar to actual Turkish *pide*. Plato mentions "cakes" made from barley flour, kneaded and cooked with olives and cheese.

Greeks familiarized the whole Mediterranean with two Egyptian procedures: the leavening and kneading of the dough to get a more digestible bread, and the use of cupola-roof-ovens instead of open fires. Greek bakers who became popular in Rome since the 4th century BC converted flavored and garnished *Plakuntos* into the roman *Placenta*=flat-bread.

The Roman poet Virgil wrote:
> *Their homely fare dispatch'd, the hungry band*
> *Invade their trenchers next, and soon devour,*
> *To mend the scanty meal, their cakes of flour.*
> *Ascanius this observ'd, and smiling said:*
> *"See, we devour the plates on which we fed."*

No doubt that more than twenty centuries ago Greeks and Romans created the prototypes of pizza, but it was the Neapolitans, "inventors" of this inexpensive food that could be consumed quickly, who were responsible for the addition of the ingredients universally associated with pizza today—tomato and mozzarella cheese (Fig. 2). Neapolitans become familiar with the "exotic" tomato plant after Columbus but the fruit was believed to be poisonous. Barely and hazardously eaten by peasants, tomatoes first appeared in the recipe book "*Il Cuoco Galante*" (The Gallant Cook) written in 1819 by chef Vincenzo Corrado. The first mentioning of "mozzarella" can be found in the recipe-book "Opera" (1570) by Bartolomeo Scappi. Up to present time, genuine mozzarella is still produced from fat buffalo milk in the vicinity of Naples (Fig. 3). It is a very delicate product, not only in taste, but also to store – mozzarella does not handle low temperatures well – if kept in a refrigerator, it becomes "rubbery". It should only be stored in its own whey at room temperature for a few days.

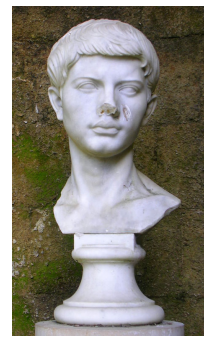

*Figure 1.* Bust of Vergil from the Tomb of Vergil in Naples, Italy

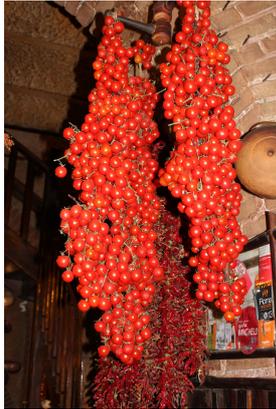
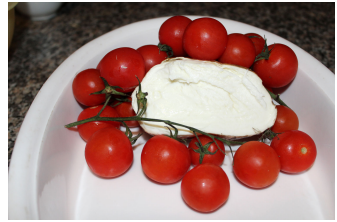

*Figure 2.* Mozzarella di Bufala and cherry tomatoes.

Documents demonstrate that until the 18th century, the Neapolitan pizza was a simple dish of pasta baked or fried, flavored with lard, pecorino-cheese, olives, salt or small fishes called *cecinielli*. During the 19th century up to two hundred "pizzaioli" (pizza-makers) crowded the streets of Naples selling for a nickel baked or fried pizzas dressed with tomato sauce and basil leaves. In 1889, a few years after the unification of Italy, the pizzaiolo Raffaele Esposito decided to pay homage to the Italian Queen by adding mozzarella to the traditional tomato and basil pie. The combination of red, white, and green symbolized the colors of the Italian flag and the tri-color pizza, still known as *Pizza Margherita*, transformed the modest pie to a success that the modest pizzaiolo could never have imagined.

In Italy today, pizza is made in different regional styles. Neapolitans are famous for their round-shaped pizzas with a high crusty edge or *cornicione*. Beyond the mentioned *Margherita* the standard is *Pizza Napoletana* (protected by the European Union as a Guaranteed Traditional Speciality), dressed with tomato, mozzarella and anchovies. The simplest version is called *Marinara*, simply topped with tomato, garlic, oregano and oil. Elaborate and opulent is the *Calzone* a circle of pizza dough folded into a half-moon stuffed with ricotta, mozzarella, salami or prosciutto.

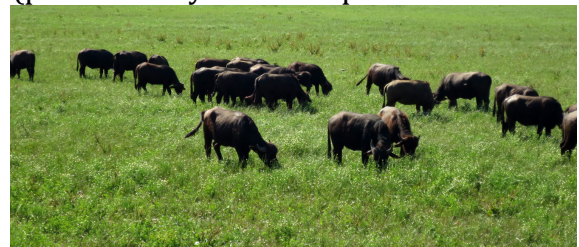

*Figure 3.* Buffalo ranch near Capaccio (Campania region), Italy

Romans, devoted to crunchy flatbreads, add more water to the flour (up to 70%) and add olive-oil or lard to the dough, so their pizzas can be stretched to the thickness of canvas without losing its toothsome chewiness; condiments are the same for the Napoletana but the pizza served at the table of a roman pizzeria is without *cornicione*. Roman bakeries and groceries usually sell *pizza bianca* by weight, a "white", rectangular, crusty pizza topped only with oil and salt. *Sardenaira* is a flatbread with tomatoes, olives & anchovies typical of Liguria but originated in the nearby Provence where it is called *pissaladière* (from the anchovy paste called *pissalat*=salt fish),

Typical of the Abruzzi's tradition is the *Pizza di Sfrigoli* made by deeply kneading lard, flour, and salt, then incorporating little pig bits (*sfrigole*) before baking, while in Apulia the beloved recipe is *Pizza Pugliese*, thin and covered with tomato sauce and a lot of stewed onions (anchovies and olives are optional). Another Apulian specialty is the

Panzerotto, a little pizza pocket served to mark the beginning of Carnival. Panzerotto differs from the Neapolitan calzone in both size and method of cooking (it's deep-fried, not baked). The classic filling is tomato sauce and fresh mozzarella, but many variations exist. Calabrians - devoted to sharp and spicy flavors - enrich their pizzas with hot salami (*soppressata*) or *'nduja*, a spreadable and fiery blend of lard and chili peppers probably introduced into Calabria by the Spanish.

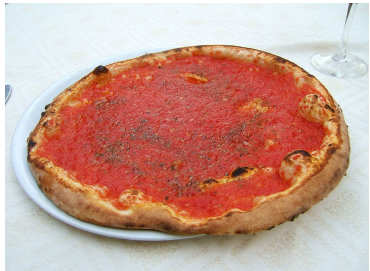

*Figure 4.* Pizza marinara

The Sicilian version of pizza is called *sfinciuni* (from the latin *spongia*=sponge), a rectangular thick and cushioned flatbread generously dressed with olive oil, onions, sheep's cheese and sun-dried tomatoes. *Scaccia* - a specialty of the Ragusa province in Sicily – is a thinly rolled dough spread with tomato sauce and cheese and then folded up on itself to resemble a strudel; the long, rectangular pizzas are then sliced, revealing the layers of crust, sauce, and cheese.

Sardinian *panada* is a nourishing pizza shield filled with eggplant, lamb and tomatoes – or in a seafood version, stuffed with fish or buttery local eel.

Essentially an "open sandwich", Neapolitan pizza disembarked to US in the late 19th and early 20th centuries when Italian immigrants, as did millions of Europeans, were coming to New York, Trenton, New Haven, Boston, Chicago and St. Louis. Flavors and aromas of humble pizzas sold on the streets by Neapolitan *pizzaioli* to the country fellows began to intrigue the Americans. Neapolitan immigrant Antonio Pero began making pizza for Lombardi's grocery store, which still exists in NY's Little Italy and in 1905 Mr. Gennaro, Lombardi's owner, was licensed to open the first Pizzeria in Manhattan on Spring Steet.  Due to the wide influence of Italian immigrants in American culture, the U.S. has developed regional forms of pizza, some bearing only a casual resemblance to the Italian original. Chicago has its own style of a deep-dish pizza. Detroit also has its unique twice-baked style, with cheese all the way to the edge of the crust, and New York City's thin crust pizzas are well-known.  St. Louis, Missouri uses thin crusts and rectangular slices in its local pizzas, while New Haven-style pizza is a thin crust variety that does not include cheese unless the customer asks for it as an additional topping.

---

**Main text:**

Being curious, the authors began looking into the secrets of making pizza. Rule number one, as Italians told them, was to always look for a pizzeria with a wood burning oven (not with an electrical one). Good pizzerias are proud of their "forno" ("oven" in Italian), in which you can see with your own eyes the whole process of baking. The *pizzaiolo* forms a dough disc, covers it with topping, places the fresh pizza on top of a wooden or aluminum spade, and finally transfers it into the oven.  A couple of minutes later it is sitting in front of you, covered with mouth-watering bubbles of cheese, encouraging you to consume it and wash it down with a pitcher of good beer.

The authors received useful advice from a friendly pizzaiolo who was working in a local Roman pizzeria, frequently visited by them when they lived in that neighborhood: "Always come for a pizza either before 8 p.m. or after 10 p.m., when the pizzeria is half-empty." The advice was also confirmed by one of the pizzeria's frequent visitors: a big grey cat. When the pizzeria was full, the cat would leave, and did not show any interest in what was on the patron's plates.

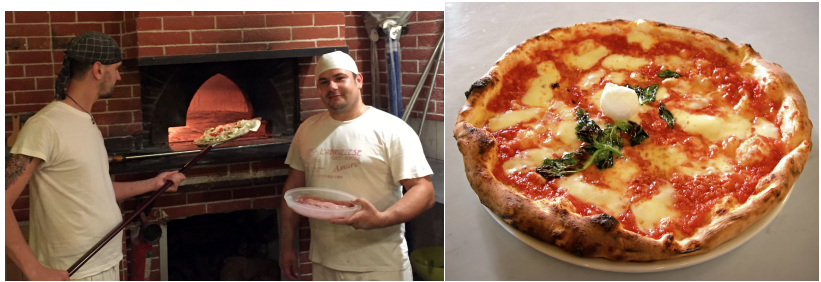

*Figure 5.* Two modern pizzaiolos in Rome in front of a brick pizza oven and pizza magherita

The reason for this advice was very simple – oven capacity. As the pizzaiolo explained, 325 - 330°C[1] is the optimal temperature for Roman pizza baked in a wood burning oven with a fire-brick bottom. In this case, a thin Roman pizza will be done in 2 minutes. Thus, even putting two pizzas into the oven, the pizzaiolo can serve 50-60 clients within an hour. During peak hours, about one hundred customers frequent the pizzeria and at least ten clients are waiting for a take-out pizza. To meet the demand, the pizzaiolo increases the temperature in the oven up to 390°C, and pizzas "fly out" of the oven every 50 seconds (hence, each one requires a "baking time" of around 1½ minutes). However, their quality is not the same: the bottom and the crust are a little "overdone" (slightly black), and the tomatoes are a little undercooked.

Since it is not always easy to find a pizzeria with a brick oven, let us take a look what advantages it has compared to an electric oven and whether there is a way to improve the latter to produce a decent pizza.

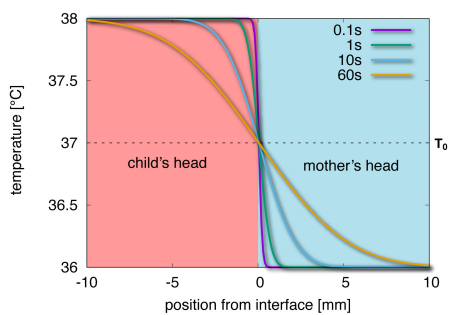

*Figure 6.* Temperature profile within mother's and child's head - 0.1s, 1s, 10s, and 60s after they made contact.

To illustrate the physical principles involved in baking pizzas, let us consider a common example of how heat is transferred. Imagine when you were a child and had a fever, but no thermometer at hand. Your mother would put her hand on your forehead and quickly say: "you have a high temperature, no school for you tomorrow". To investigate this process scientifically, we start with simplifying the problem. Let us imagine that your mom is touching your forehead with her own forehead rather than her hand. In that case, if the temperature of your forehead would have been 38°C, and your mother's 36°C, it is clear by the symmetry of the problem that the temperature at the interface ($T_0$) between the two foreheads will be 37°C, and that your mother would feel the flow of heat coming from your forehead (the actual temperature distribution in time is shown in Fig. 6).

---

[1] Obviously, the temperature depends on the way the dough was prepared and stored. The pizzaiolo Antonio prepares the dough well in advance – 24 hours before baking the pizza. After mixing all ingredients and kneading it well, he leaves the dough "to rest" for several hours, and then cuts it into pieces and forms ball-shaped portions. For Naples-style pizza the portion is 180-250 grams, for Roman-style less. These portions are used for a single pizza. Then he puts the dough balls into wooden boxes, where the dough will be rise for 4-6 hours. After that it is ready to be baked or placed in a refrigerator for later use.

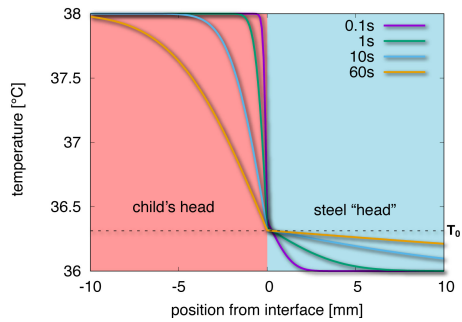

*Figure 7.* The same as Fig. 6, but with a cooler steel "head".

Now let us assume that your mother's head is made of steel, and her temperature is the same – 36°C. Intuitively, it is clear the temperature at the interface will decrease, let us say, to 36.3°C. This is related to the fact that the steel will draw off the heat from the interface region effectively to its bulk, since its heat conductivity is large. It is also clear that this removal becomes more efficient, when a smaller amount of heat needs to be drawn away from the interface region (*i.e.*, the heat removal increases when the specific heat capacity of the "mother's head" material decreases, an illustration is shown in Fig. 7).

Let us now analyze the process of pizza baking more scientifically. We start by reminding the reader of the main concepts of heat transfer [1]. When we speak about "heat", we usually have in mind the energy of a system (like the mother's head, the oven, or the pizza itself) associated with the chaotic motion of atoms, molecules and other particles it is composed of. We inherited this concept of heat from the physics of a past era. Physicists say that heat is not a function of the state of a system, its amount depends on the way the system achieved this state. Like work, heat is not a type of energy, but rather a value convenient to use to describe energy transfer [2,3]. The amount of heat, necessary to raise the temperature of a mass unit of the material by one degree, is called a specific heat capacity of the material:

$$c = \frac{\Delta Q}{M \Delta T} \qquad (1)$$

Here M is the mass of the system and $\Delta Q$ is the quantity of a heat required for heating the system by a temperature $\Delta T$. From this expression it is clear that the heat capacity is measured in J×kg$^{-1}$×K$^{-1}$ in SI units.

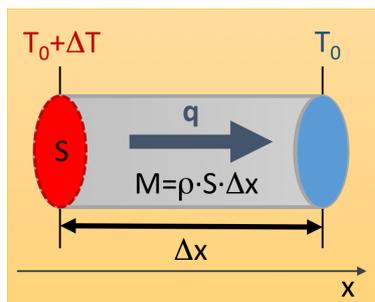

*Figure 8.* Heat flow in a small cylinder from hot ($T_0+\Delta T$) to cold ($T_0$). Notice, the temperature decreases from left to right!

In the case of a thermal contact between the two systems with different temperatures, the heat will go from the warmer system to the cooler one. The heat flux density $q$ is the amount of heat $\Delta Q$ that flows through a unit area per unit time in the direction of temperature change:

$$q = \frac{\Delta Q}{S \Delta t} \qquad (2)$$

In the simplest case of a homogenous non-uniformly heated system, using Eq. (1), one finds

$$q = \frac{c M \Delta T}{S \Delta t} = c\rho \frac{(\Delta x)^2}{\Delta t}\left(\frac{\Delta T}{\Delta x}\right) = -\kappa \frac{dT}{dx}, \qquad (3)$$

where $\rho$ is the mass density[2]. Assuming that $\Delta x$ is small, we identified the value in parentheses as the derivative of the temperature by the coordinate x and took into account the fact that the temperature decreases in x-direction (see Fig. 8). In the general case, **q** is the vector and the derivative in Eq. (3) is replaced by the gradient $\boldsymbol{\nabla} T$, which describes the rate of temperature change in space. The coefficient $\kappa$ in Eq. (3) is the thermal conductivity, which describes the ability of a material to transfer heat when a temperature gradient is applied[3]. Eq. (3) expresses mathematically the so-called Fourier's law, which is valid when the temperature variation is small.

---

[2] One can also use this formula to easily find the heat loss through the walls of ones house during a cold winter. In this stationary situation, the temperature distribution does not change with time.

Next, let us analyze how a "temperature front" penetrates a medium from its surface, when a heat flow is supplied to it (see Fig. 8). Assume that during time t the temperature in the small cylinder of the height $L(t)$ and cross-section $S$ has changed by $\Delta T$. [4] Let us get back to Eq. (3) and rewrite it by replacing $\Delta x$ by $L(t)$:

$$\frac{c\rho L(t)\Delta T}{t} = \kappa \frac{\Delta T}{L(t)}. \tag{4}$$

Solving Eq. (4) with respect to the length $L(t)$ one finds:

$$L(t) \sim \sqrt{\frac{\kappa t}{c\rho}} = \sqrt{\chi t}, \tag{5}$$

i.e., the temperature front enters the medium by the square-root law of time. The time after which the temperature at depth L will reach a value close to the one of the interface depends on the values $\kappa$, c, and $\rho$. The parameter $\chi = \kappa/c\rho$ is called the thermal diffusivity or coefficient of temperature conductivity and the heating time of the whole volume can be expressed in its terms: $\tau \sim L^2/\chi$.

Of course, our consideration of the heat penetration problem into a medium is just a simple evaluation of the value $L(t)$. A more precise approach requires solution of differential equations. Yet, the final result confirms our conclusion (5), just corrected by a numerical factor:

$$L(t) = \sqrt{\pi \chi t}. \tag{6}$$

Now that we know how heat transfer works, let us get back to the problem of calculating the temperature at the interface between two semi-spaces: on the left with parameters $\kappa_1$, $c_1$, $\rho_1$ and temperature T₁ at $-\infty$, and on the right with parameters $\kappa_2$, $c_2$, $\rho_2$ and temperature T₂ at $+\infty$. Let us denote the temperature at the boundary layer as T₀. The equation of the energy balance, i.e., the requirement of the equality of the heat flowing from the *warm*, right semi-space through the interface to the *cold,* left semi-space, can be written in the form

$$q = \kappa_1 \frac{T_1 - T_0}{\sqrt{\pi \chi_1 t}} = \kappa_2 \frac{T_0 - T_2}{\sqrt{\pi \chi_2 t}}. \tag{7}$$

Here we simplified the problem assuming that all temperature changes happen at the corresponding temperature dependent length (6). Solving this equation with respect to $T_0$ one finds that

$$T_0 = \frac{T_1 + \nu_{21} T_2}{1 + \nu_{21}}, \tag{8}$$

where

$$\nu_{21} = \frac{\kappa_2}{\kappa_1} \sqrt{\frac{\chi_1}{\chi_2}} = \sqrt{\frac{\kappa_2 c_2 \rho_2}{\kappa_1 c_1 \rho_1}}. \tag{9}$$

---

[3] The definition of the thermal conductivity $\kappa = c\rho \frac{(\Delta x)^2}{\Delta t}$ used in Eq. (3) requires a clarification: While our simplified derivation suggests a geometry dependence, we emphasize that in reality it is determined only by microscopic properties of the material.

[4] This process is not stationary anymore and the flow q not constant, since the heat will be partially used for the heating of the cylinder material. Therefore, unlike in the stationary process, the rate of temperature change dT/dx in the medium is a function of space and time.

One notices, that time does not enter in expression (8) (*i.e.*, the interface temperature remains constant in the process of the heat transfer, see Figs. 6,7&9). In the case of identical media with different temperatures one can easily find: $T_0 = \frac{T_1+T_2}{2}$. This is the quantitative proof of the intuitive response we provided in the beginning of the article for the temperature 37°C of the interface between the mother's hand and the child's forehead. If the mother's hand would be made of steel, $\nu_{21} \gg 1$ and $T_0 \approx T_2$, its temperature would remain almost unchanged after contact with the hot forehead, meaning that she would not be able to notice the child's fever.

Finally, we are ready to discuss the advantages of the brick oven. Let us start from the calculation of the temperature at the interface between the pizza placed into the brick oven and its heated baking surface. All necessary parameters are shown in the table below:

| Property / Material | Heat capacity c [J/(kg×K)] | Thermal conductivity κ [W/(m×K)] | Mass density ρ [kg/m³] | Temperature conductivity χ [m²/s] | $\nu_{21}$ [5] |
|---|---|---|---|---|---|
| dough[6] | 2-2.5×10³ | 0.5 | 0.6-0.8×10³ | 2.5-4.2×10⁻⁷ | 1 |
| food grade steel (X18H10T) | 4.96×10² | 18 | 7.9×10³ | 4.5×10⁻⁶ | 0.1 |
| fire brick | 8.8×10² | 0.86 | 2.5×10³ | 4.0×10⁻⁷ | 0.65 |
| Water (@25C) | 4.2×10³ | 0.58 | 1.0×10³ | 1.4×10⁻⁷ | 0.2 |

*Table 1.* Physical properties of different materials, including heat capacity, thermal conductivity, density and temperature conductivity

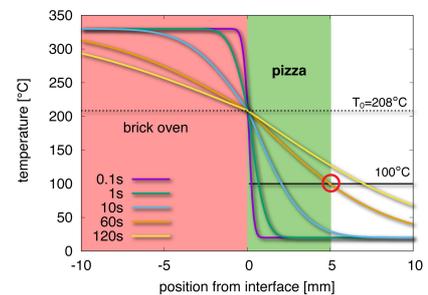

*Figure 9.* Temperature profile in a brick oven with pizza at different times. At 60s the top surface of the pizza reaches 100°C (red circle). Here we only take thermal diffusion into account. Evaporation and radiation are neglected.

Assuming the initial temperature of the pizza dough as $T_0^{do}$ =20°C, and the temperature inside the oven, as our pizzaiolo claimed, being about $T_1^{wo}$ =330°C, we find for the temperature at the boundary layer between the oven surface and pizza bottom

$$T_0^{wo} = \frac{330\ ^0C + 0.65 \cdot 20\ ^0C}{1.65} \approx 208\ ^0C.$$

As we know from the words of the same pizzaiolo, a pizza is perfectly baked in two minutes under these conditions.

Let us now repeat our calculations for the electric oven with its baking surface made of steel. For an electric oven the ratio will be $\nu_{eo} = 0.1$, and if heated to the same temperature of 330°C, the temperature at the bottom of the pizza will be equal to

$$\frac{330\ ^0C + 0.1 \cdot 20\ ^0C}{1.1} \approx 300\ ^0C.$$

That is too much! The pizza will just turn into coal! This interface temperature is even much higher than in Naples' pizzerias, where oven temperatures between 400-450°C are common.

---
[5] For dough, steel, and brick, material "2" is dough. For water, material "1" is steel.
[6] The material parameters for dough should be considered as an estimate. It is clear that the exact numbers strongly depend on the type of flour and the fermentation/rising time of the dough (in the latter process, the dough is enriched by gases, which change its density).

Well, let us formulate the problem differently. Let us assume that generations of pizza makers, who were using wooden spade to transfer pizzas into the oven, are right: the temperature at the (Roman) pizza's bottom should be about 210°C. What would be the necessary temperature for an electric oven with steel surface?

The answer follows from the Eq. (8) with coefficient $v_{eo} = 0.1$ and solved with respect to $T_1^{eo}$ when the temperature at the bottom of the pizza is the same as in the wood oven: $T_0^{eo} = T_0^{wo}$. The result of this exercise shows that the electric oven should be much colder than the brick one: $T_1^{eo} \approx 230\ ^{\circ}C$.

It seems that if one is able to forget the flavor of burning wood, sizzling dry air in the brick oven and the other natural features, the problem would be solved – let us set electric stove controls to 230°C and in a couple of minutes we can take an excellent pizza out of the oven. But: Is it that easy?

In order to answer this question, we first need to consider the second important mechanism of a heat transfer: *thermal radiation* [4]. Its intensity, the amount of radiation energy arriving each second to 1cm² of surface in the oven, is determined by the Stefan-Boltzmann law:
$$I = \sigma T^4, \tag{10}$$
where $\sigma = 5.67 \cdot 10^{-8}\ W/(m^2 K^4)$ is the so-called Stefan-Boltzmann constant.

A typical brick oven has a double-crown vault filled with sand, which is kept at almost constant temperature. Its walls as well as the bottom part, are also heated to $T_1^{wo} = 330°C = 603K$, meaning that the complete volume of the oven is "filled" by infrared radiation. With a temperature that high, this radiation becomes significant: the pizza here is continuously "irradiated" from both sides by a "flow" of infrared radiation of the intensity
$$I^{wo} = \sigma(T_1^{wo})^4 = 5.67 \cdot 10^{-8} (603)^4 = 7.5 kW/m^2,$$
i.e., each second an amount of energy close to the 0.75 J arrives at 1cm² of pizza.[7]

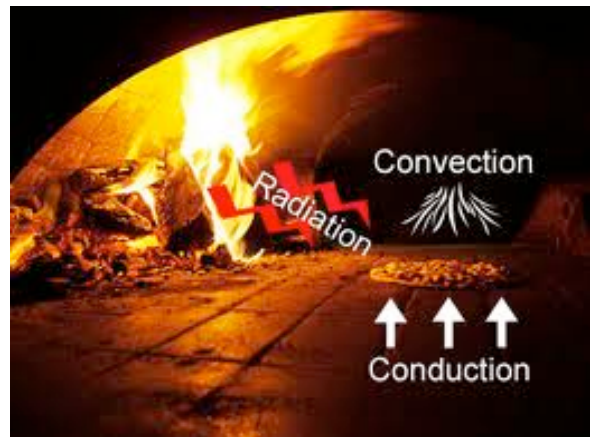

*Figure 10.* Heat transfer mechanisms in the pizza oven.

Here one should notice, that, in its turn, the pizza also irradiates out a "flow" of the intensity $I_{pizza} = \sigma(T_{pizza})^4$. Since the major part of the baking time is required for the evaporation of water contained in the dough and toppings, we can assume $T_{pizza} = T_b = 100°C = 373\ K$, which results in a radiation intensity of $I_{pizza} = \sigma(T_b)^4 = 1.1\ kW/m^2$, i.e., 15% of the obtained radiation, the pizza "returns" to the oven.

For the much less heated electric oven, the corresponding amount of energy, incident to 1cm² of pizza surface, is more than twice less:
$$I^{eo} = \sigma(T_1^{eo})^4 = 5.67 \cdot 10^{-8} \frac{(503)^4 W}{m^2} = 3.6 kW/m^2,$$

---

[7] Here we assume that the pizza behaves as a black body. In reality it is slightly reflective, reducing the amount of heat it absorbs.

while the returned radiation is the same: $1.1 kW/m^2$.

Now it is a time to evaluate what amount of heat 1 cm² of pizza receives per second through its bottom. By the definition it is determined by the heat flow (3) and in order to get its numeric value we will evaluate the temperature gradient at the oven surface in the same way as was already done in Eq. (7):
$$q(t) = \kappa \frac{T_1^o - T_0}{\sqrt{\pi \chi t}},$$
where $T_1^o$ is the temperature of the oven. One can see, that, contrary to the Stefan-Boltzmann radiation, the heat flux arriving into the pizza by means of a heat conductance depends on time. Correspondingly, the amount of heat transferred to 1 cm² of pizza in this way from the oven during time $\tau$ is determined by
$$Q(\tau) = \int_0^\tau q(t) dt = 2\kappa(T_1^o - T_0)\sqrt{\frac{\tau}{\pi \chi}}.$$
Therefore, the total amount of heat, arriving at 1 cm² of pizza during time $\tau$, is
$$Q_{tot}(\tau) = \sigma[(T_1^o)^4 - (T_b)^4]\tau + 2\kappa(T_1^o - T_0)\sqrt{\frac{\tau}{\pi\chi}}. \qquad (11)$$
This value is used to heat 1 cm² of pizza from the dough temperature $T_2^d = 20°C$ to the boiling temperature of water $T^b = 100°C$:
$$Q_{heat} = c^d \rho^d d(T^b - T_2^d).$$
Yet, this is not all. During the process of baking the perfect pizza we apparently evaporate water from the dough, tomatoes, cheese, and other ingredients. We need to take the required energy for this into account as well. If one assumes that the water mass fraction α evaporates from the dough and all topping one gets:
$$Q_{boil} = \alpha L \rho^{water} d.$$
Here $d$ is the thickness of the pizza, which we assume to be $d$=0.5 cm, while $L = 2264.76 \, J \cdot g^{-1}$ is the latent heat of evaporation for water.
Collecting both these contributions in one, we can write
$$Q_{tot} = Q_{heat} + Q_{boil} = c^{do}\rho^{do}d(T^b - T_2^d) + \alpha L \rho^{water} d. \qquad (12)$$

Equating Eqs. (11) and (12) one finds the final equation determining the "baking time" of pizza:
$$\sigma[(T_1^o)^4 - (T^b)^4]\tau + 2\kappa(T_1^o - T_0)\sqrt{\frac{\tau}{\pi\chi}} = c^{do}\rho^{do}d(T^b - T_2^d) + \alpha L \rho^{water} d. \qquad (13)$$

In order to obtain a realistic answer for the baking time, it is important to know the amount of water, which is evaporated during the baking process. A typical recipe for pizza Margarita calls for 240g of dough and 90g of toppings (consisting of tomatoes and mozzarella). The dough contains about one-third of water and the toppings 80% (the rest is mostly fat from the cheese). Together with a weight loss of 30g, a good assumption is a 20% loss of water, i.e. α=0.2. Using this with the values of specific heat capacity and density for dough from the table above, one finds that $Q_{tot} = (70 + 226) J/cm^2$, which gives for the baking time in the wood oven $\tau_{wo} \approx 125$ s. For the electric oven an analogous calculation results in an almost 50% longer time $\tau_{eo} \approx 170s$. We see that we have succeeded to reproduce the value disclosed to us by our pizzaiolo: 2 minutes for baking in a wood oven. The result of an attempt to bake a pizza in the electric oven will be the mentioned unbalanced product.

Using Eq. (8) one can easily find that the temperature at the interface between the pizza and oven surface reaches 240°C, when the temperature in wood fired brick oven increases to 390°C. Replacing correspondingly $T_0$ in Eq. (13) one can find the baking time under these extreme conditions to be approximately 82 seconds, hence the productivity of the oven increases by almost 50%!

A final "trick" disclosed to us is the importance for pizzas with water-rich toppings (eggplants, tomatoes slices, or other vegetables). In this case, the expert first bakes the pizza in the regular way on the oven surface, but when the pizza's bottom is ready he lifts it with the wooden/aluminum spade and holds it elevated from the baking surface for another half minute or more in order to expose the pizza to just heat irradiation. In this way they avoid burning the dough and get well cooked toppings.

Certainly, as is routinely done in physics, in order to get to the core of the phenomenon, we examined only the simplest model here (in particular, we ignored the third mechanism of the heat transfer: convection, which we can assume to have only a small effect. See Fig.~10), which captures the essential physical processes.

As a final note, we remark that it is difficult to build a classic brick oven, and many customers do not appreciate the difference between an excellent and decent pizza. These are the reasons why engineers invent all sorts of innovations: for example, inserting a ceramic bottom made of special ceramics to imitate the bottom of brick ovens in a modern professional electric oven. To bake a pizza evenly, rotating baking surfaces are available – convection ovens emulate the gas flows in brick ovens, and many other things. But, the dry heat and the smell of wood in traditional firebrick ovens remain the ideal way to bake the perfect pizza.

*Acknowledgements*

We want to express our gratitude to the Roman pizzaiolos Antonio and Vincenzo for lifting the veil to some secrets of the pizza baking art. Special thanks goes to Mr. Zheng Zhou, a young student of a high school in Shanghai, attending the lectures of one of the authors (AV) for making several valuable comments, which helped us to improve the final text of this article.